\documentclass[aps,prd,twocolumn,superscriptaddress,nofootinbib,floatfix]{revtex4-2}

\usepackage{graphicx}   
\usepackage{amssymb}    
\usepackage{amsmath}   
\usepackage{bm}
\usepackage{xcolor}
\usepackage[colorlinks=true,allcolors=blue]{hyperref}

\newcommand{\bz}{b(z)}
\newcommand{\mz}{M(z)}
\newcommand{\de}{\mathrm d}
\newcommand{\g}{$\gamma$}
\newcommand{\be}{\begin{equation}}
\newcommand{\ee}{\end{equation}}

\begin{document}

\title{On the Clustering Bias of Unresolved Gamma-Ray Sources}

\author{Bhashin A. Thakore}
\email{bhashinashish.thakore@unito.it}
\affiliation{Dipartimento di Fisica, Universit\`a degli Studi di Torino, via P. Giuria 1, 10125 Torino, Italy}
\affiliation{INFN -- Istituto Nazionale di Fisica Nucleare, Sezione di Torino, via P. Giuria 1, 10125 Torino, Italy}
\affiliation{GRAPPA (Gravitation Astroparticle Physics Amsterdam), University of Amsterdam, Science Park 904, 1098 XH Amsterdam, The Netherlands}

\author{Marco Regis}
\affiliation{Dipartimento di Fisica, Universit\`a degli Studi di Torino, via P. Giuria 1, 10125 Torino, Italy}
\affiliation{INFN -- Istituto Nazionale di Fisica Nucleare, Sezione di Torino, via P. Giuria 1, 10125 Torino, Italy}

\author{Michela Negro}
\affiliation{Department of Physics \& Astronomy, Louisiana State University, Baton Rouge, LA 70803, USA}

\author{Stefano Camera}
\affiliation{Dipartimento di Fisica, Universit\`a degli Studi di Torino, via P. Giuria 1, 10125 Torino, Italy}
\affiliation{INFN -- Istituto Nazionale di Fisica Nucleare, Sezione di Torino, via P. Giuria 1, 10125 Torino, Italy}
\affiliation{INAF -- Istituto Nazionale di Astrofisica, Osservatorio Astrofisico di Torino, strada Osservatorio 20, 10025 Pino Torinese, Italy}
\affiliation{Department of Physics \& Astronomy, University of the Western Cape, Cape Town 7535, South Africa}

\author{Daniel Gruen}
\affiliation{University Observatory, Faculty of Physics, Ludwig-Maximilians-Universität, Scheinerstr. 1, 81679 Munich, Germany}
\affiliation{Excellence Cluster ORIGINS, Boltzmannstr. 2, 85748 Garching, Germany}

\author{Nicolao Fornengo}
\affiliation{Dipartimento di Fisica, Università degli Studi di Torino, via P. Giuria 1, 10125 Torino, Italy}
\affiliation{INFN -- Istituto Nazionale di Fisica Nucleare, Sezione di Torino, via P. Giuria 1, 10125 Torino, Italy}

\author{Aaron J. Roodman}
\affiliation{SLAC National Accelerator Laboratory, Menlo Park, CA 94025, USA}
\affiliation{Kavli Institute for Particle Astrophysics \& Cosmology, P.O. Box 2450, Stanford University, Stanford, CA 94305, USA}

\begin{abstract}

The Unresolved \g-Ray Background (UGRB) encodes the collective emission of source populations that are too faint to be detected individually, and its cross-correlation with tracers of the large-scale structure has emerged as a powerful tool to characterize those populations. In this work, we analyze the weak-lensing--UGRB cross-correlation using a reference blazar model. By modeling the relation between blazar $\gamma$-ray luminosity and host-halo mass, we infer the average linear bias $\langle b(z) \rangle$ and host-halo mass $\langle M(z) \rangle$ of the unresolved blazar population. We find that the signal is dominated by moderately biased sources, with $\langle b \rangle \gtrsim 2$ at $z\simeq 1$, hosted by halos with $\langle M \rangle \sim \mathcal{O}(10^{13}\,M_\odot)$. Repeating the analysis with a misaligned-AGN model yields consistent bias values, indicating that the result is determined by the clustering properties of the UGRB rather than by the details of the $\gamma$-ray source model.
\end{abstract}

\maketitle

\section{Introduction}
\label{sec:intro}

As unique messengers of the high-energy Universe, \g-rays provide insight into some of its most energetic phenomena. Because they carry information across vast spatial scales and cosmic epochs, \g-rays probe both astrophysics and fundamental physics, including indirect searches for dark matter. Individually resolved objects, however, cannot account for all of the observed extragalactic emission. After known sources and the Galactic foreground are removed, the remaining emission, i.e, the  remaining UGRB is generally attributed to the cumulative glow of cosmic populations too faint to be detected individually.
Determining which populations make up the UGRB, and their relative contributions, is a long-standing problem. Blazars (dominant class of $\gamma$-ray resolved sources), misaligned active galactic nuclei (mAGNs), star-forming galaxies (SFGs), and possibly more exotic sources such as dark-matter (DM) annihilation are all expected to contribute~\cite{ajello2015origin}. A powerful method for separating these contributions is to cross-correlate UGRB maps with tracers of the large-scale structure of the Universe \cite{camera2013novel,Fornengo2014}, such as the clustering of galaxies and galaxy clusters \cite{ando2014mapping,xia2015tomography,regis2015particle,cuoco2015dark,shirasaki2015cross,cuoco2017tomographic,Ammazzalorso:2018evf,paopiamsap2024constraints,branchini2017cross,hashimoto2019measurement,colavincenzo2020searching,tan2020bounds,Pinetti:2025hgd,krolewski2026tomography}, weak lensing \cite{camera2015tomographic,shirasaki2014cross,troster2017cross,shirasaki2016cosmological,shirasaki2018correlation,DES:2019ucp,Zhang:2026ysp}, and cosmic microwave background lensing \cite{fornengo2015evidence,feng2017planck}. These observables trace the distribution of matter across cosmological distances and thus encode complementary information about the origin of \g-ray emission. Since the various contributors differ in spectral shape, angular separation and
redshift evolution, each leaves a distinct imprint on the cross-correlation, thereby allowing one to disentangle their individual contributions.
In two recent studies, we measured the cross-correlation of the UGRB with Dark Energy Survey (DES) Year 3 data. First, with the tangential shear of weakly lensed source galaxies \cite{thakore2025high}
\footnote{A similar analysis was recently presented in Ref.~\cite{shirasaki2026constraints}. We repeated their statistical approach and obtained compatible results. On the other hand, in Ref.~\cite{thakore2026multi} (Appendix A) we reported a higher statistical significance for the cross-correlation signal. This is mainly because we fitted a (well-motivated) model to data
and compared it against the null signal. This procedure allows us to properly weigh bins depending on the expected signal. In Ref.~\cite{shirasaki2026constraints}, by contrast, bins where no signal is expected are also included in the $\chi^2$ computation, which increases the number of degrees of freedom and correspondingly lowers the significance.}, then with the projected overdensity of \texttt{redMaGiC} luminous red galaxies, combining the two probes in a multi-tracer analysis \cite{thakore2026multi}. In both cases the detection is driven by the two-halo term, so the measurements are primarily sensitive to the \textit{large-scale clustering} of the unresolved \g-ray sources rather than to their occupation of individual host halos. 

Fitting the cross-correlation data with a reference blazar model, we found that an unusually large normalization is required to match the measurements, corresponding to a rescaling factor of $5.4^{+1.9}_{-1.6}$~\cite{thakore2026multi}.
This excess may be due partly to a larger number density of blazars (an increase by a factor of $\sim 2$ in the $\gamma$-ray luminosity function (GLF) relative to the reference model remains compatible with $\gamma$-ray number counts and angular auto-correlation measurements(Refs.~\cite{Ackermann:2018wlo,korsmeier2022flat}) and partly to a larger bias than that adopted in Refs.~\cite{thakore2025high,thakore2026multi}. The latter would imply that unresolved blazars reside in more massive halos than assumed in the baseline model.

In this work, we attempt to disentangle the latter contribution and to estimate the clustering bias of the blazars that populate the UGRB.
To this end, we start from the well-established bias of halos with respect to matter~\cite{Cooray2002}.
Generalizing the relation used in Refs.~\cite{camera2015tomographic,DES:2019ucp,thakore2025high,thakore2026multi}, we define a phenomenological relation between blazar luminosity at a given redshift and the mass of its host halo. We assume a power-law luminosity scaling described by two free parameters, a normalization and a slope, which are determined by the fit. The theoretically predicted halo bias then determines the blazar bias.

From the statistical analysis of the cross-correlation measurement, we derive the flux-weighted linear bias $\langle b(z)\rangle$ and average host-halo mass $\langle M(z)\rangle$ of the blazars that occupy the unresolved $\gamma$-ray source population, thereby improving the characterization of the UGRB.

\section{Data and methodology}
\label{sec:data}

This work builds upon recent measurements of cross-correlations between the UGRB and tracers of large-scale cosmic structure (see, e.g., Ref.~\cite{thakore2026multi}). The \g-ray sky is probed with twelve years of \textit{Fermi}-LAT observations, from which the UGRB component is extracted. The resulting flux maps are divided into nine broad, logarithmically spaced energy bins spanning $631\,\mathrm{MeV}$ to $1\,\mathrm{TeV}$. The matter distribution is traced with three years of Dark Energy Survey (DES Y3) data through the tangential shear of weakly lensed source galaxies in the first study, and through the projected overdensity of \texttt{redMaGiC} luminous red galaxies (galaxy clustering) in the second. The cross-correlations are measured in configuration space across twelve angular bins between $5$ and $600\,\mathrm{arcmin}$. A detailed description of the datasets and covariance estimation is provided in Ref.~\cite{thakore2026multi}. Here, the measurement is interpreted with a physical halo-model description in which blazars (BLZ) dominate the \g-ray source population contributing to the detected cross-correlation signal.

The measurements in Refs.~\cite{thakore2025high,thakore2026multi} detect positive lensing--UGRB and galaxy--UGRB correlations, with most of the significance arising from large angular scales. The lensing result shows that a substantial component of the UGRB traces the matter distribution measured by weak lensing; its statistical significance is driven by the two-halo term rather than by a small population of extremely bright, nearby objects. The multi-tracer analysis finds the single-tracer galaxy--UGRB and lensing--UGRB results to be mutually consistent. Their combination yields a signal-to-noise ratio of 10.3, providing strong evidence that the majority of the UGRB component is extragalactic and traces the large-scale structure.

\begin{table}[!t]
    {\centering
    \begin{tabular}{lll}
    \hline
    Parameter & Prior & Median (68\% C.I.) \\
    \hline
    $\log_{10}A_{\rm BLZ,eff}$ & $\mathcal{U}(-10,\,2)$   & $0.50^{+0.23}_{-0.20}$ \\
    $\mu_{\rm BLZ}$            & $\mathcal{U}(1.5,\,3.0)$  & $1.87^{+0.16}_{-0.20}$ \\
    $p_1$                      & $\mathcal{U}(1.0,\,8.0)$  & unconstrained \\
    $\log_{10}(M_0/M_\odot)$   & $\mathcal{U}(13,\,15)$    & $14.09^{+0.71}_{-0.52}$ \\
    $\alpha$                   & $\mathcal{U}(0,\,1)$      & $0.25^{+0.12}_{-0.25}$ \\
    \hline
    \end{tabular}
    }
    \caption{Priors adopted in the analysis, and the corresponding marginalized medians
    with 68\% credible intervals for the lensing--UGRB cross-correlation. The amplitude is
    quoted as sampled, in $\log_{10}$, corresponding to
    $A_{\rm BLZ,eff} = 3.16^{+2.20}_{-1.16}$.}
    \label{tab:priors_lensing}
\end{table}

\section{Halo occupation of blazars and clustering bias}
\label{sec:bias}
The cross-correlation signal can be computed from the angular power spectrum involving \g-ray\ sources and the matter tracer as~\cite{Fornengo2014}
\be
 C_\ell^{ar} = \int \, \frac{\de E\, \de z}{H(z)} 
 \frac{W_\textrm{\g}^a(E,z)\, W_{\rm t}^r(z)}{\chi^2(z)} 
 \,P_{\gamma{\rm t}}\left[k=\frac\ell{\chi(z)}, z\right]\;,
 \label{eq:clgen}
\ee
where $\chi(z)$ is the comoving distance at redshift $z$, satisfying $\de z / \de \chi = H(z)$ in a flat Universe, with $H(z)$ being the Hubble parameter. 
$W_\gamma^a(E,z)$ and $W_{\rm t}^r(z)$ are the window functions for energy bin $a$ and redshift bin $r$, respectively, and describe the redshift distributions of the $\gamma$-ray emission and the matter tracer. The cross-power spectrum $P_{\gamma {\rm t}}$ is the sum of its one- and two-halo components, $P_{\gamma {\rm t}}^{\rm 1h}$ and $P_{\gamma {\rm t}}^{\rm 2h}$. Within the Limber approximation, the comoving wavenumber $k$ and angular multipole $\ell$ are related by $k = \ell / \chi(z)$.
The cosmological parameters entering Eq.~\eqref{eq:clgen} are adopted from Ref.~\cite{DES:2021wwk}.

The \g-ray window function of Eq.~(\ref{eq:clgen}) is given by~\cite{DES:2019ucp}
\begin{equation}
\begin{split}
W_\textrm{\g}^a(E,z)=\chi^2(z)\,\int_{\Gamma_{\rm min}}^{\Gamma_{\rm max}} \de \Gamma\,
\int_{\mathcal{L}_{\rm min}}^{\mathcal{L}_{\rm max}}& \de \mathcal{L}\,
 \Phi_{\rm S}(\mathcal{L},z,\Gamma)\,\frac{\de F}{\de E}\,,
\end{split}
\label{eq:window}
\end{equation}
with $\de F/\de E \propto E^{-\Gamma}\,e^{-\tau}$ being the differential photon spectrum, $\tau$ the optical depth due to absorption on the extragalactic background light~\cite{Finke2010}, and $\mathcal{L}_{\rm max}(\Gamma)$ ensuring that only unresolved sources are included. We use $\mathcal{L}_{\rm min}=7\times10^{43}$ erg/s (Refs.~\cite{ajello2014cosmic,korsmeier2022flat}), $\Gamma_{\rm min}=1.4$ and $\Gamma_{\rm max}=2.8$ (exact values are not very relevant for our results). The gamma-ray luminosity function (GLF) is $\Phi_{\rm S}(\mathcal{L},z,\Gamma)\equiv \frac{\de ^3n}{\de V\de \mathcal{L}\de \Gamma}$, i.e., the number of sources per unit luminosity $\mathcal{L}$, comoving volume $V$ at redshift $z$, and photon spectral index $\Gamma$.

The one- and two-halo components of the three-dimensional cross-power spectrum between the matter density and astrophysical $\gamma$-ray sources are given by
\begin{align}
     P_{\gamma_{\rm S}\delta}^{\rm 1h}(k,z) &=\int_{\Gamma_{\rm min}}^{\Gamma_{\rm max}} \de \Gamma\, \int_{\mathcal{L}_{\rm min}}^{\mathcal{L}_{\rm max}} \de \mathcal{L}\,\frac{\Phi_{\rm S}(\mathcal{L},z,\Gamma)}{\langle f_{\rm S} \rangle}\nonumber \\
     &\times \frac{\de F}{\de E}\left(\mathcal{L},z,\Gamma\right)\, \hat u_\delta\left(k|M(\mathcal{L},z),z\right)\nonumber\\
     P_{\gamma_{\rm S}\delta}^{\rm 2h}(k,z) &= \left[\int_{\Gamma_{\rm min}}^{\Gamma_{\rm max}} \de \Gamma\,\int_{\mathcal{L}_{\rm min}}^{\mathcal{L}_{\rm max}} \de \mathcal{L}\, b_{\rm S}(\mathcal{L},z)\,\frac{\Phi_{\rm S}(\mathcal{L},z,\Gamma)}{\langle f_{\rm S} \rangle} \frac{\de F}{\de E} \right]\nonumber \\
     &\times \left[\int_{M_{\rm min}}^{M_{\rm max}} \de M\,\frac{\de n}{\de M} b_{\rm h}(M,z) \hat u_\delta(k|M,z) \right] 
      P^{\rm lin}(k,z) ,
	\label{eq:PSastro}
\end{align}
where $u_\delta(k|M,z)$ is the Fourier transform of the halo density profile and $b_{\rm S}$ is the bias of \g-ray astrophysical sources with respect to the matter density, for which we adopt $b_{\rm S}(\mathcal{L},z)=b_{\rm h}[M(\mathcal{L},z)]$. Thus, a source with luminosity $\mathcal{L}$ has the same bias $b_{\rm h}$ as a halo of mass $M(\mathcal{L},z)$; the relation between host-halo mass and source luminosity is described below.
The mean flux is defined as $\langle f_{\rm S} \rangle=\int \de \mathcal{L}\, \de \Gamma\,\de F/\de E\, \Phi_{\rm S}$.
Because the inferred bias depends on the assumed $\gamma$-ray source model, we summarize its explicit form here.  We decompose the GLF into its local form at $z=0$ and an evolutionary factor $e(\mathcal{L},z)$,
\be
\Phi_{\rm S}(\mathcal{L},z,\Gamma)=\Phi_{\rm S}(\mathcal{L},0,\Gamma)\times e(\mathcal{L},z)\,,
\ee
where $\mathcal{L}$ is the rest-frame luminosity in the $(0.1-100)$~GeV range. At $z=0$ the GLF reads~\cite{korsmeier2022flat}
\begin{equation}
\begin{split}
\Phi_{\rm S}(\mathcal{L},0,\Gamma)=&\frac{A}{\ln(10)\,\mathcal{L}}
\left[(\mathcal{L}/\mathcal{L}_0)^{\kappa_1}+(\mathcal{L}/\mathcal{L}_0)^{\kappa_2}\right]^{-1} \\
&\times \exp\left[-\frac{(\Gamma-\mu_{\rm BLZ})^2}{2\sigma^2}\right]\,,
\end{split}
\label{eq:glf0}
\end{equation}
where $A$ is the normalization constant, $\mathcal{L}_0$ the break luminosity, $\kappa_1$ and $\kappa_2$ describing the luminosity dependence of the GLF, and the Gaussian term accounting for the distribution of spectral indices around a mean value $\mu_{\rm BLZ}$ with dispersion $\sigma$. The redshift evolution is parametrized as
\be
e(\mathcal{L},z)=\left[\left(\frac{1+z}{1+z_c}\right)^{-p_1}+\left(\frac{1+z}{1+z_c}\right)^{-p_2}\right]^{-1}\,,
\label{eq:evol}
\ee
where $p_1$ and $p_2$ set the growth and the decline of the source density below and above the peak redshift $z_c$, and both $z_c$ and $\mu_{\rm BLZ}$ depend on $\mathcal{L}$~\cite{korsmeier2022flat}. 

\begin{figure}[!t]
    \centering
    \includegraphics[width=\linewidth]{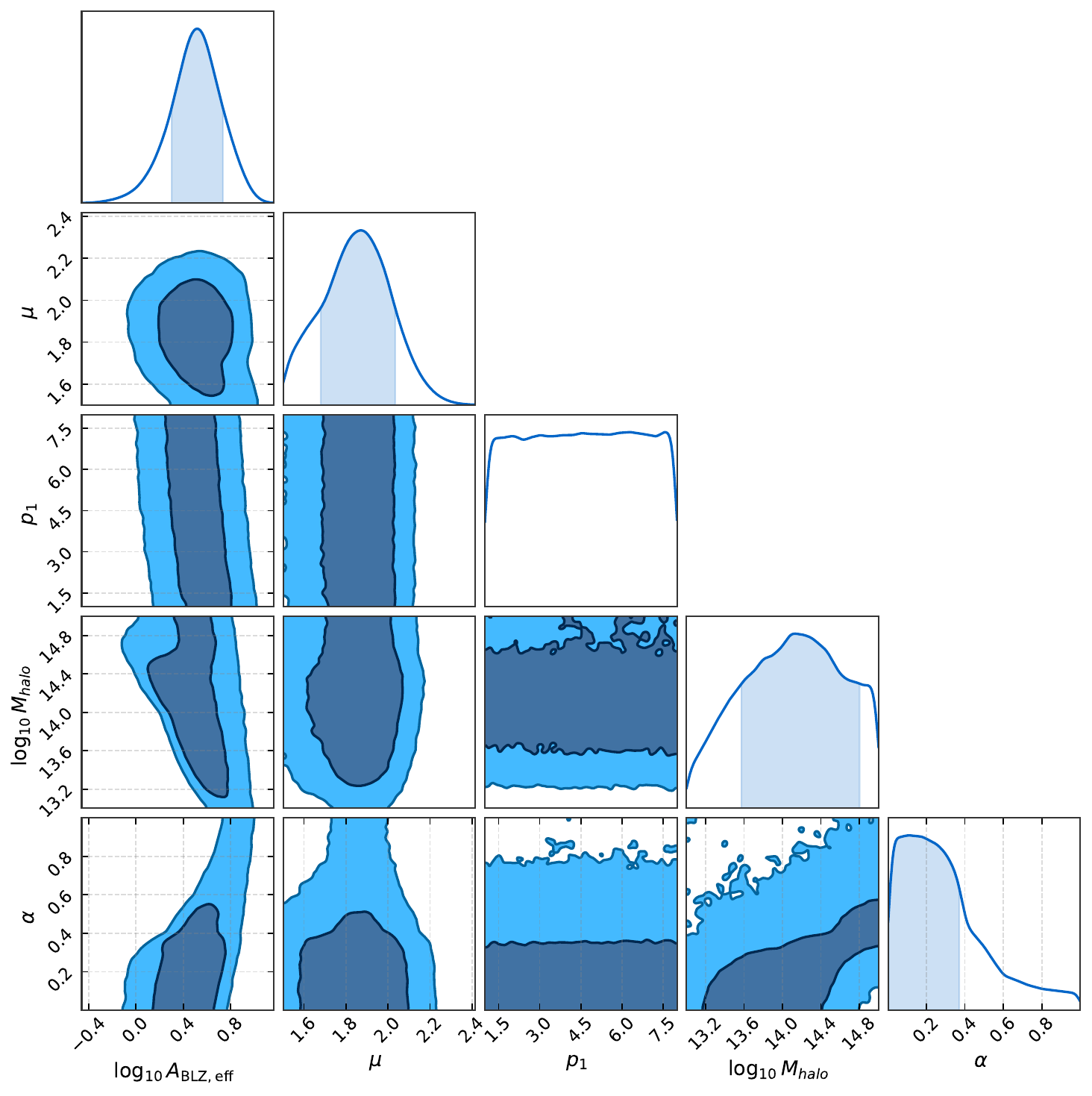}
    \caption{Posterior distributions of the parameters of the blazar model, for the lensing-UGRB cross-correlation. The 2D contours refer to the
    68\% and 95\% credible regions, while the shaded areas in the 1D subplots denote the
    68\% credible interval.}
    \label{fig:posterior_lensing}
\end{figure}

Eq.~\eqref{eq:PSastro} shows that the mass--luminosity relation affects the one- and two-halo terms differently. Its effect on the cross-power spectrum can therefore, in principle, be separated from an overall amplitude that rescales the GLF relative to the reference model.

This approach requires a reliable estimate of the one-halo term, i.e., of the occupation of the matter tracer by blazars. Because weak lensing directly traces the matter distribution, Eq.~\eqref{eq:PSastro} can be applied without an additional tracer-occupation model. For a galaxy tracer, by contrast, one must model how blazars populate the specific galaxy sample. This relation is difficult to establish for the color- and luminosity-selected \texttt{redMaGiC} sample, in particular because BLZ are associated to star formation processes, whilst the \texttt{redMaGiC} sample is made of red galaxies. 
Consequently, although the cross-correlation of \textit{Fermi}-LAT $\gamma$ rays with \texttt{redMaGiC} galaxies is an important probe of the cosmological origin of the UGRB, it is not a clean observable for this bias determination. We therefore restrict the present analysis to the weak-lensing cross-correlation.

We now describe our model for the relation between blazar $\gamma$-ray luminosity and host-halo mass. In Ref.~\cite{camera2015tomographic}, this relation was obtained by associating the $\gamma$-ray luminosity with the mass of the supermassive black hole powering the AGN and then relating the black-hole mass to the DM-halo mass:
\begin{equation}
M(\mathcal{L},z)=2\times 10^{13}M_\odot\left[\mathcal{L}/(10^{47}\,\mathrm{erg\,s^{-1}})\right]^{0.23}(1+z)^{-0.9}\,.
\end{equation}

\begin{figure*}[htbp!]
    \centering
        \includegraphics[width=\textwidth]{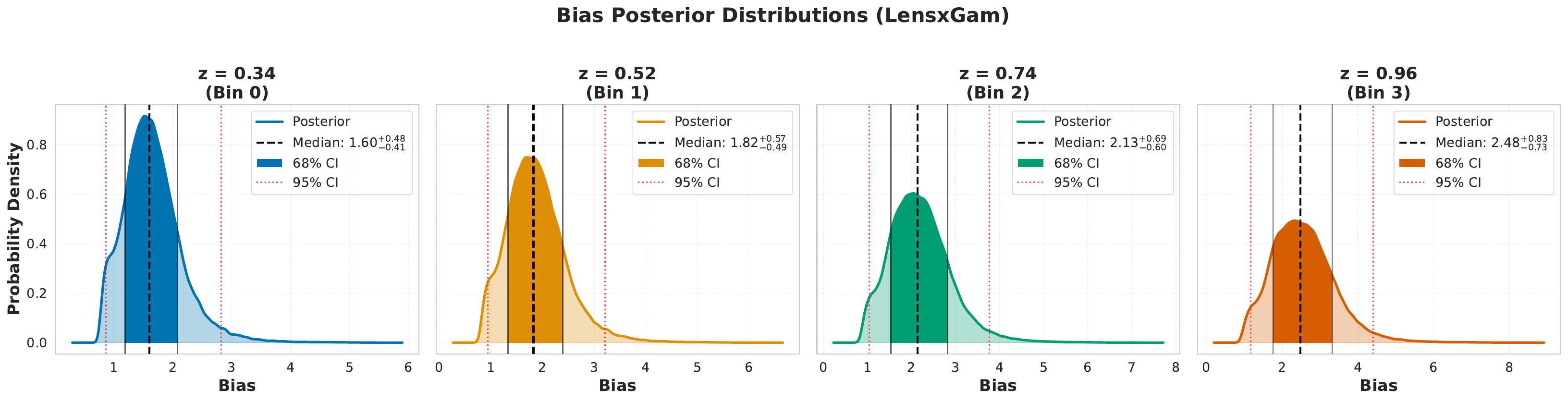}
        \caption{Posterior distributions for the flux-averaged bias $\langle\bz\rangle$ for the lensing-UGRB
    cross-correlation, in the four DES~Y3 source-galaxy redshift bins.  The dashes vertical lines denotes the median values, while the shaded regions and the dotted lines denote the 68\% and 95\% credible interval.}
        \label{fig:bias_posteriors_lensing}
\end{figure*}

\begin{figure*}[!t]
    \centering
        \includegraphics[width=\linewidth]{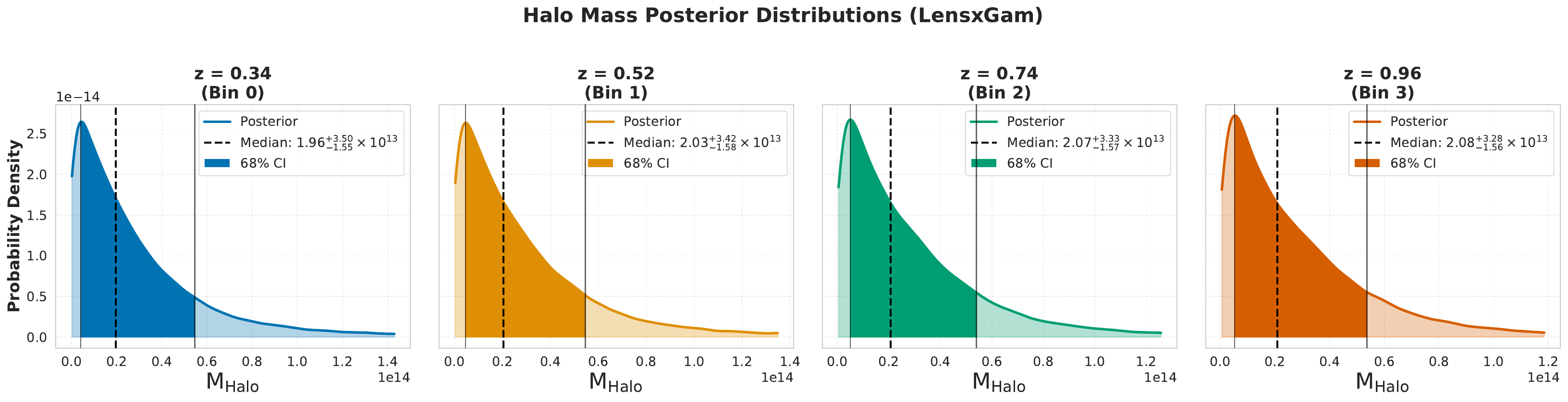}
        \caption{Posterior distributions for the flux-averaged host-halo mass $\langle M(z)\rangle$ for the lensing-UGRB
    cross-correlation, in the four DES~Y3 source-galaxy redshift bins.  The dashes vertical lines denotes the median values, while the shaded regions and the dotted lines denote the 68\% and 95\% credible interval.}
        \label{fig:Mhalo_posteriors_lensing}
\end{figure*}

Here we adopt a more general $M(\mathcal{L},z)$ relation:
\begin{equation}
M(\mathcal{L},z) = M_0\left(\mathcal{L}/10^{47}\,\mathrm{erg\,s^{-1}}\right)^{\alpha}(1+z)^{-0.9}\,,
\end{equation}
with $M_0$ and $\alpha$ as free parameters constrained by the statistical analysis.\footnote{The redshift dependence of the cross-correlation is not strongly constrained by our measurement (see Ref.~\cite{thakore2025high} and the following), so the redshift index is kept fixed for simplicity.}

Then, we derive the flux-averaged bias of \g-ray astrophysical sources with respect to the matter density as
\begin{equation}
\begin{split}
\langle b(z) \rangle = \int_{\Gamma_{\rm min}}^{\Gamma_{\rm max}} \de \Gamma\,\int_{\mathcal{L}_{\rm min}}^{\mathcal{L}_{\rm max}} \de \mathcal{L}
 & \, b_{\rm h}[M(\mathcal{L},z)]\,
\frac{\Phi_{\rm S}(\mathcal{L},z,\Gamma)}{\langle f_{\rm S} \rangle} \\
& \times \frac{\de F}{\de E}\left(\mathcal{L},z,\Gamma\right) \,.
\end{split}
\label{eq:bastro}
\end{equation}
Here $b_{\rm h}$ is the halo bias, and a source of luminosity $\mathcal{L}$ is assigned the bias of its host halo of mass $M$.

Similarly, we compute the average mass of halos hosting blazars as
\begin{equation}
\begin{split}
\langle M(z) \rangle=
\int_{\Gamma_{\rm min}}^{\Gamma_{\rm max}} \de \Gamma\,\int_{\mathcal{L}_{\rm min}}^{\mathcal{L}_{\rm max}} \de \mathcal{L} & M(\mathcal{L},z)\,\frac{\Phi_{\rm S}(\mathcal{L},z,\Gamma)}{\langle f_{\rm S} \rangle} \\
& \times \frac{\de F}{\de E}\left(\mathcal{L},z,\Gamma\right) \,.
\end{split}
\label{eq:aveM}
\end{equation}
In addition to $M_0$ and $\alpha$ in the mass--luminosity relation and the overall amplitude $A_{\rm BLZ,eff}$ in Eq.~\eqref{eq:physmdl}, the free parameters are the spectral-index parameter $\mu_{\rm BLZ}$ in Eq.~\eqref{eq:glf0} and the redshift-evolution parameter $p_1$ in Eq.~\eqref{eq:evol}. All the remaining GLF parameters are fixed to the best-fit values for BL Lacs derived from \g-ray number counts and angular auto-correlation in Ref.~\cite{korsmeier2022flat} (the BLL 4FGL$+C_P$ fit of their Table~2). In particular, $\kappa_1$ is degenerate with $A$ and, as shown in Ref.~\cite{thakore2026multi}, varying it within its uncertainty is reabsorbed in the overall normalization. We therefore keep it fixed.

\begin{figure}[htbp!]
    \centering
      \includegraphics[width=\columnwidth]{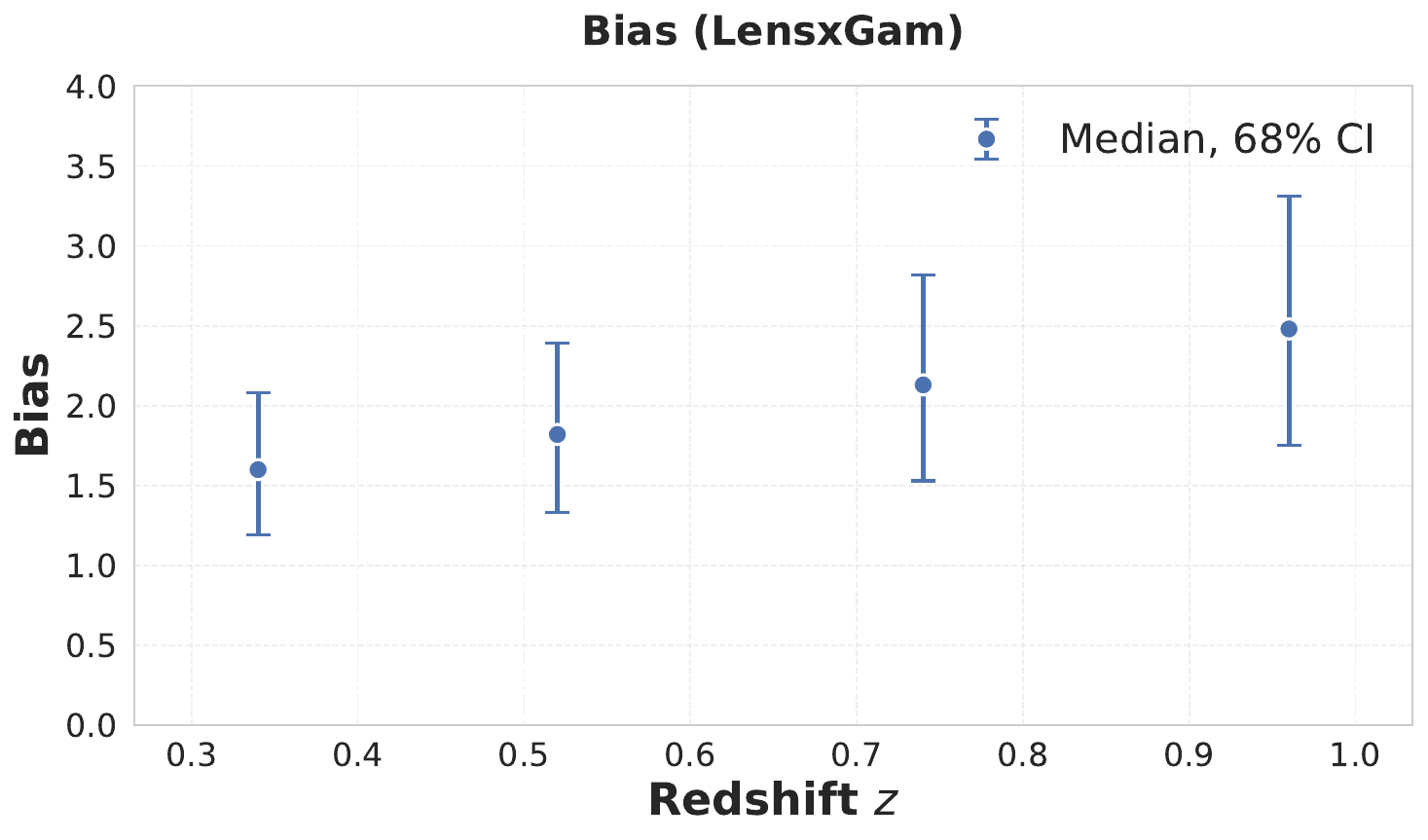}
        \includegraphics[width=\columnwidth]{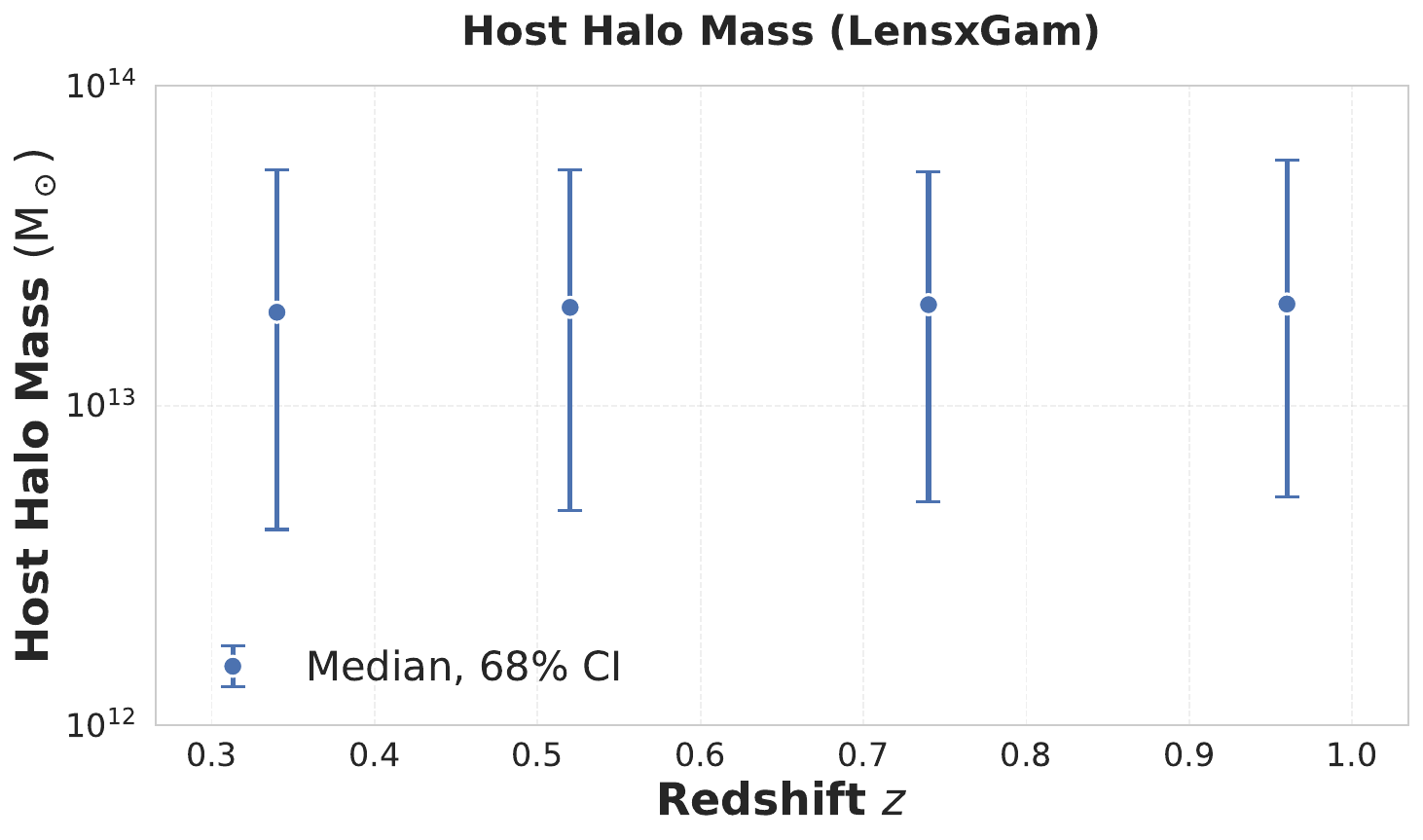}
        \caption{Median values and 68\% credible intervals of the flux-averaged bias $\langle\bz\rangle$ (top panel) and flux-averaged host-halo mass $\langle M(z)\rangle$ (bottom panel), in the four DES energy bins of our analysis.}
        \label{fig:errorbars_lensing}
\end{figure}

\section{Results}
\label{sec:results}

To measure the flux-averaged bias and host-halo mass in Eqs.~\eqref{eq:bastro} and~\eqref{eq:aveM}, we compare the predicted blazar contribution with the measured angular cross-correlation function. As mentioned above, the blazar model has five free parameters: the spectral-index parameter $\mu_{\rm BLZ}$ and redshift-evolution parameter $p_1$ entering the GLF, the mass scale $M_0$ and luminosity scaling $\alpha$ entering the $M(\mathcal{L},z)$ relation, and an overall normalization $A_{\rm BLZ,eff}$. The blazar cross-correlation function is therefore
\begin{align}
\Xi_{\rm phys}^{ar}(\theta) &=
A_{\rm BLZ,eff}\, \Xi_{\rm BLZ}^{ar}(\theta, \mu_{\rm BLZ}, p_1, M_0, \alpha) \,.
\label{eq:physmdl}
\end{align}
$A_{\rm BLZ,eff}$ is an effective blazar amplitude common to the one- and two-halo terms, unlike the two independent normalizations used in Refs.~\cite{thakore2025high,thakore2026multi}. It can therefore be interpreted as a rescaling of the GLF, while the freedom in the mass--luminosity relation modifies the relative one- and two-halo contributions.

We infer the model parameters using Markov chain Monte Carlo (MCMC) sampling with the affine-invariant ensemble sampler implemented in \texttt{emcee}~\cite{foreman2013emcee}. The posterior distributions are visualized with \texttt{ChainConsumer}~\cite{hinton2016chainconsumer}. The priors adopted for the five physical parameters are listed in Table~\ref{tab:priors_lensing}.

The posterior distributions obtained from the lensing-UGRB cross-correlation are shown in Fig.~\ref{fig:posterior_lensing}. The amplitude has a posterior median of $A_{\rm BLZ,eff} = 3.2^{+2.2}_{-1.2}$. The posterior of $\log_{10}(M_0/M_\odot)$ peaks at $\sim 14.1$, significantly above the $M_0 = 2\times10^{13}M_\odot$ of the reference relation in Ref.~\cite{camera2015tomographic}. The median spectral index $\mu_{\rm BLZ} = 1.87^{+0.16}_{-0.2}$ is broadly consistent with the values obtained in Refs.~\cite{thakore2025high, thakore2026multi}.

Since $A_{\rm BLZ,eff}$ and $M_0$ are anticorrelated in the posterior (Fig.~\ref{fig:posterior_lensing}), a larger host mass, and hence a larger bias,  accounts for part of the normalization of the signal, reducing $A_{\rm BLZ,eff}$. Comparing with the two-halo amplitude obtained from the same data in Ref.~\cite{thakore2026multi}, the median falls from $A^{\rm 2h}_{\rm BLZ} = 5.4^{+1.9}_{-1.6}$ to $A_{\rm BLZ,eff} = 3.2^{+2.2}_{-1.2}$, and the best fit from $A^{\rm 2h}_{\rm BLZ} = 4.2$ to $A_{\rm BLZ,eff} = 2.2$. The shift is in the direction expected if the excess large-scale signal reported previously is partially driven by an underestimated linear bias.

Using Eqs.~\eqref{eq:bastro} and~\eqref{eq:aveM}, we derive the posterior distributions and central 68\% credible intervals of the flux-averaged bias $\bz$ and host-halo mass $\mz$, shown in Figs.~\ref{fig:bias_posteriors_lensing}, \ref{fig:Mhalo_posteriors_lensing} and \ref{fig:errorbars_lensing}.  The median bias grows steadily across the four DES~Y3 source bins, from $\langle \bz\rangle = 1.60^{+0.48}_{-0.41}$ at $\langle z \rangle = 0.34$ to  $\langle \bz\rangle = 2.48^{+0.83}_{-0.73}$ at $\langle z \rangle = 0.96$. In the third and fourth bins, which contain most of the signal~\cite{thakore2025high}, the median bias is $\gtrsim 2$. The host masses, by contrast, are statistically indistinguishable across the four bins, with medians confined to $\mz \sim (1.96\text{-}2.08)\times10^{13}M_\odot$ and 68\% intervals spanning $\sim(0.1\text{-}0.5)\times10^{14}M_\odot$. The rise in bias occurs at essentially constant host mass and is therefore consistent with the redshift growth of the halo bias function, rather than with an evolving blazar host population. The mass posteriors are broad and positively skewed, favoring host halos of a few times $10^{13}M_\odot$ within the 68\% credible intervals.

We can also compare the values we obtained with the clustering measurement of resolved
\textit{Fermi}-LAT 2LAC blazars in Ref.~\cite{allevato2014clustering}, which found
BL~Lacs and FSRQs to reside in halos of $\log_{10}(M_{\rm h}/h^{-1}M_\odot)\simeq 13.35$
and $13.40$ at $z\simeq0.4$ and $z\simeq1.2$, respectively, i.e.$M_{\rm h}\simeq(3\text{-}4)\times10^{13}M_\odot$. Our determination for the unresolved population lies slightly below, but still within, this range at the 68\% level, suggesting that faint, unresolved blazars occupy environments similar to those of their resolved counterparts.

\section{Conclusion}
\label{sec:conclusions}
In this paper, we determine the clustering bias and average host-halo mass of UGRB sources. Building on our previous UGRB cross-correlation measurements~\cite{thakore2025high,thakore2026multi}, we focus on blazars; the case of misaligned AGNs is considered in the Appendix. We introduce a phenomenological relation between blazar luminosity and host-halo mass and constrain its free parameters, which in turn determine the bias.

The sources are found to be moderately biased, with $\langle b\rangle\gtrsim 2$ in the high-redshift bins that dominate the signal, corresponding to hosts of $\langle M\rangle\sim 2\times 10^{13}M_\odot$.

In Refs.~\cite{thakore2025high,thakore2026multi}, we found the $\gamma$-ray--lensing cross-correlation to be dominated by the two-halo term (large-scale clustering), with a large required amplitude $A^{\rm 2h}_{\rm BLZ}$. Allowing the bias to vary reduces the best-fit amplitude from $A^{\rm 2h}_{\rm BLZ}\sim 4.2$ to $A_{\rm BLZ,eff}\sim 2.2$.
We therefore infer a larger bias, and hence a larger average host-halo mass, than assumed in Refs.~\cite{thakore2025high,thakore2026multi}; this substantially reduces the rescaling required by the reference model.
A modification of the GLF, to be investigated elsewhere, could account for the remaining factor of $\sim 2$ while remaining compatible with number-count and auto-correlation measurements~\cite{korsmeier2022flat}. The resulting blazar model therefore provides a more consistent description of the UGRB fluctuations. 

Angular cross-correlations are a powerful tool for determining the properties of cosmological $\gamma$-ray source populations, and the improved statistics of forthcoming lensing and galaxy surveys will enable more detailed studies of the UGRB.

\begin{acknowledgments}
MR, BT and NF acknowledge support from the Research grant TAsP (Theoretical Astroparticle Physics) funded by \textsc{infn}. BT acknowledges the support provided by the PNRR grant ex DM 118 scholarship.The work of MR and NF  is supported by the European Union -- Next Generation EU and by the Italian Ministry of University and Research (MUR) via the PRIN 2022 project n.\ 20228WHTYC -- CUP D53C24003550006.
\end{acknowledgments}

\begin{figure}[!t]
    \centering
    \includegraphics[width=\linewidth]{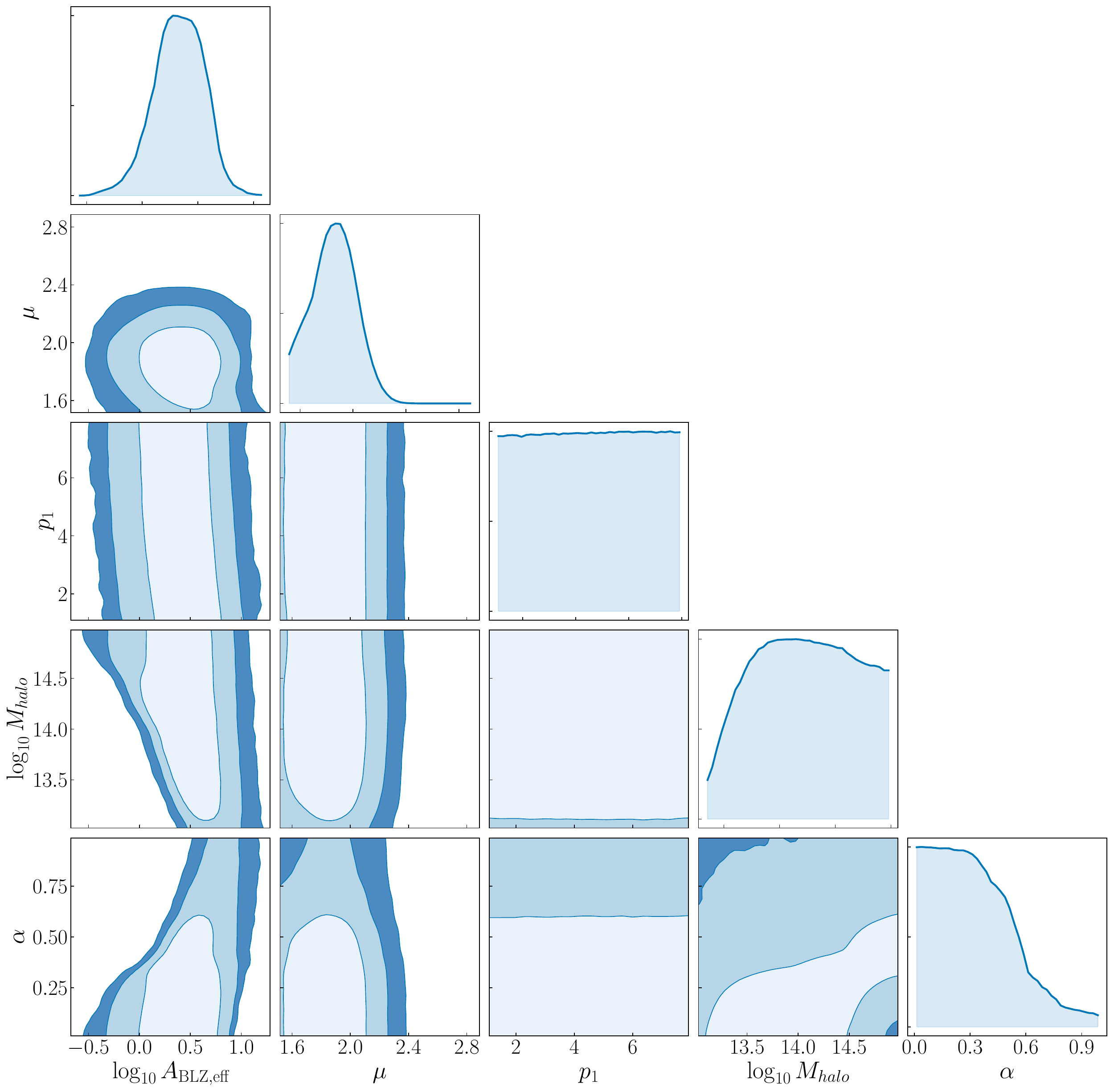}
    \caption{Similar to Fig.~\ref{fig:posterior_lensing} but showing the profile likelihood contours of the model parameters instead of their posterior distributions.}
    \label{fig:profile_likelihood_lensing}
\end{figure}

\appendix

\section{Frequentist statistical analysis}
\label{sec:prof_like}

To verify that the results of Sec.~\ref{sec:results} are not driven by posterior-volume effects, we repeat the analysis with a frequentist approach, profiling the likelihood over the parameters not shown rather than marginalizing over them. The resulting profile likelihoods are shown in Fig.~\ref{fig:profile_likelihood_lensing}. The degeneracy directions are the same as in Fig.~\ref{fig:posterior_lensing}, and the best-fitting regions are consistent with the marginalized posterior constraints.

\section{Misaligned AGNs}
\label{sec:mAGN}

\begin{figure}[!t]
    \centering
    \includegraphics[width=\linewidth]{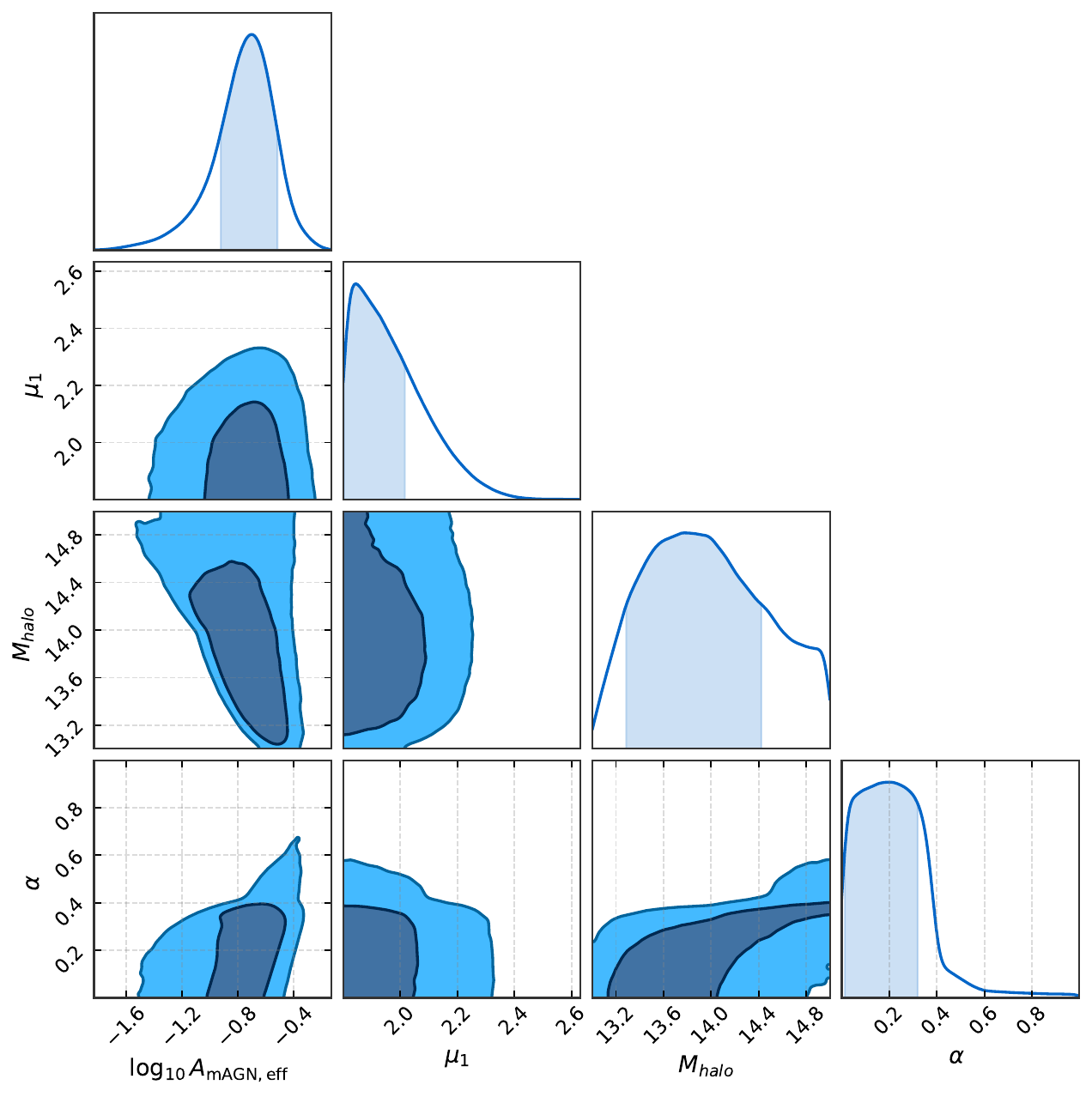}
    \caption{Posterior distributions of the parameters of the mAGN model. The 2D contours refer to the 68\% and 95\% credible regions, while the shaded areas in the 1D subplots denote the 68\% credible interval.}
    \label{fig:posterior_mAGN_lensing}
\end{figure}

\begin{figure*}[!t]
    \centering
        \includegraphics[width=\linewidth]{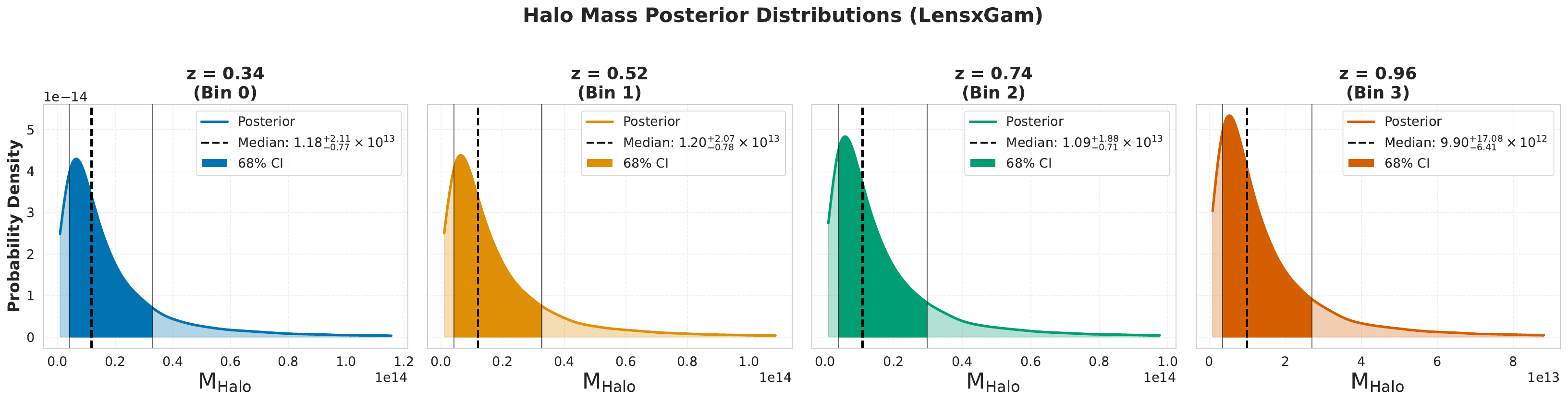}
        \caption{As in Fig. \ref{fig:Mhalo_posteriors_lensing}, but for mAGNs.}
        \label{fig:Mhalo_posteriors_lensing_mAGN}
\end{figure*}

\begin{figure}[!t]
    \centering
      \includegraphics[width=\columnwidth]{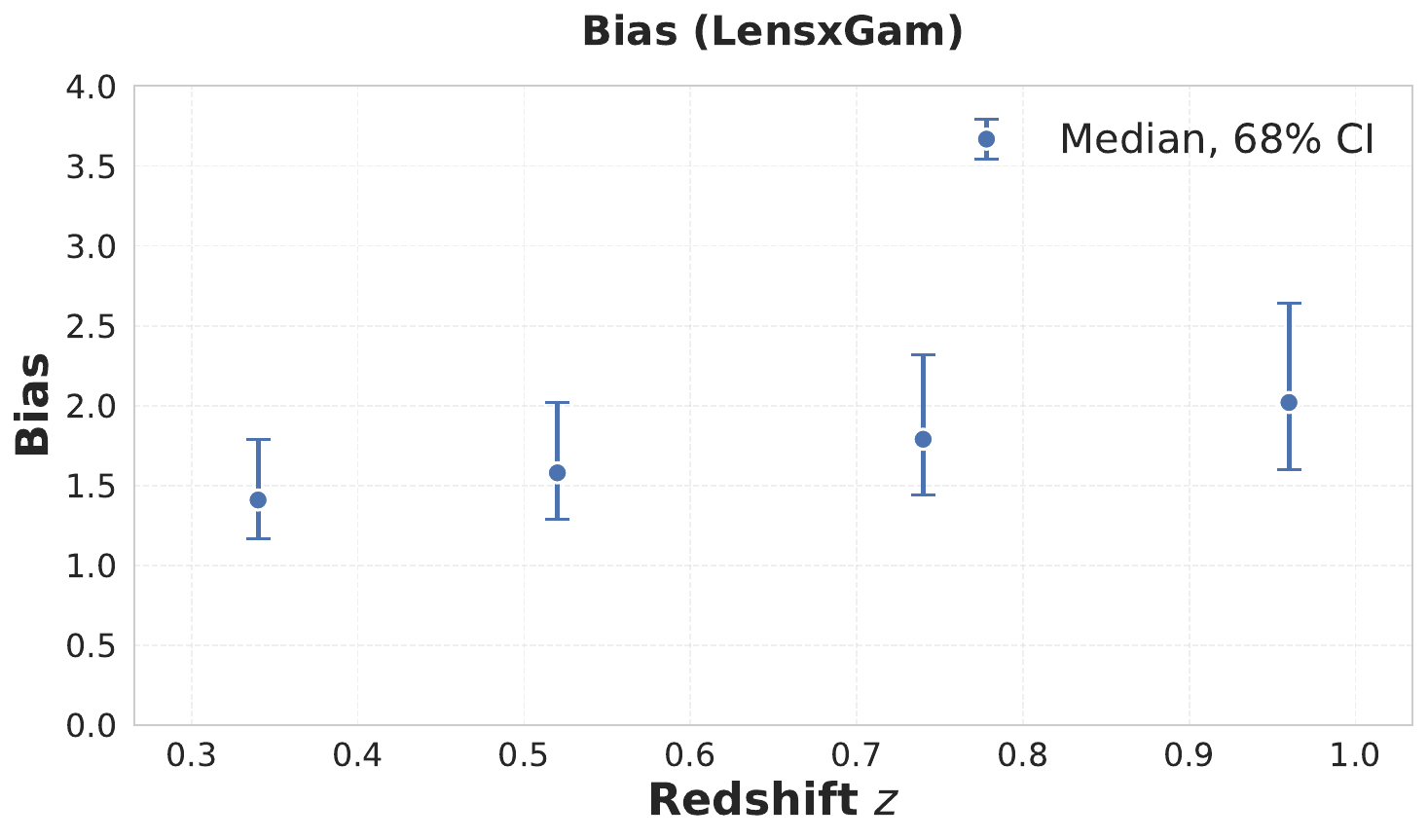}
        \includegraphics[width=\columnwidth]{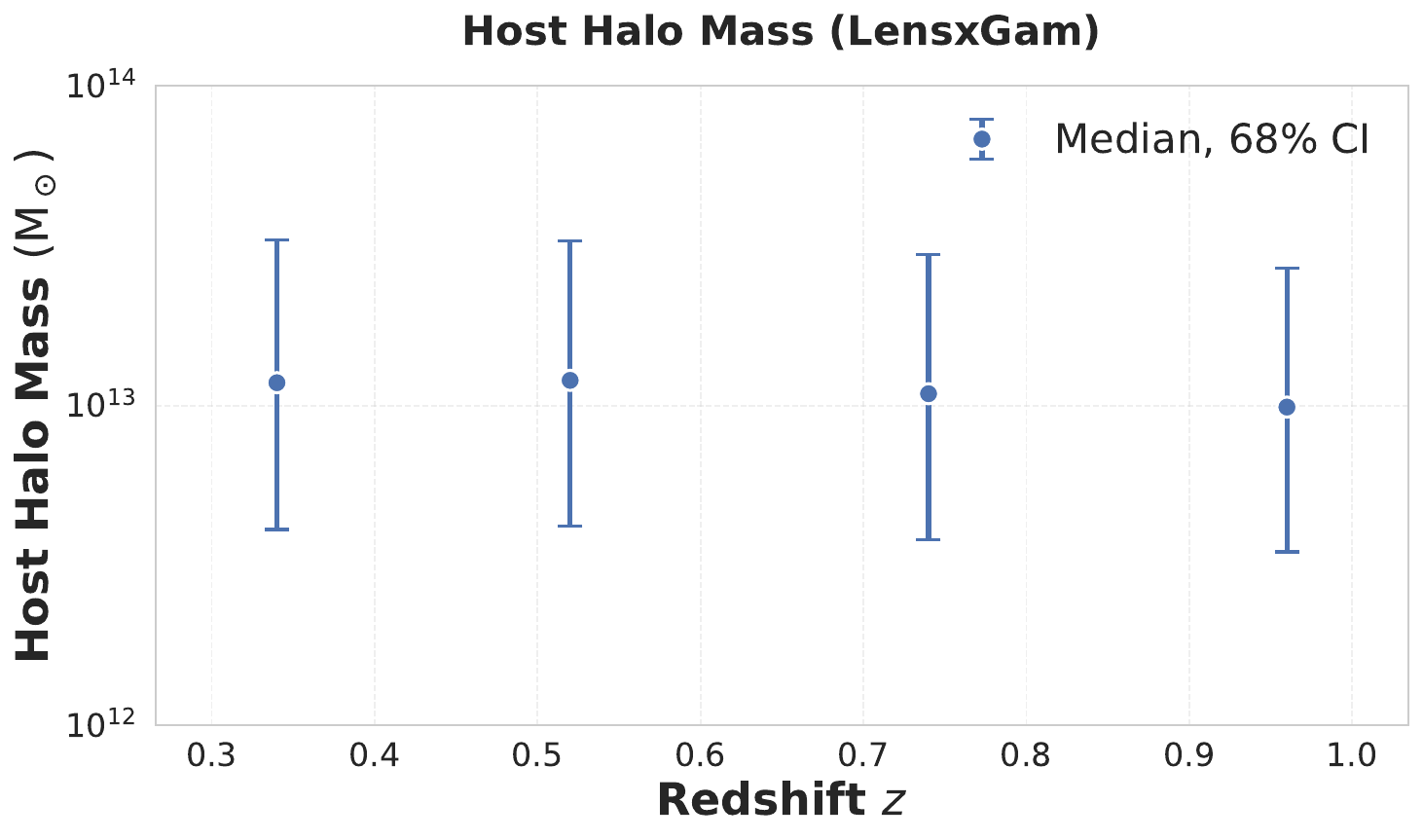}
        \caption{As in Fig. \ref{fig:errorbars_lensing}, but for mAGNs.}
        \label{fig:errorbars_lensing_mAGN}
\end{figure}

We adopt blazars as our baseline population because they provide a good fit to several UGRB probes~\cite{korsmeier2022flat,thakore2025high}; in particular, the hardness of the measured $\gamma$-ray energy spectrum disfavors populations with softer spectra. It is nevertheless useful to ask how strongly the bias determination of Sec.~\ref{sec:results} depends on this choice. We therefore repeat the analysis with $M_0$ and $\alpha$ free, replacing the blazar GLF with a model of misaligned AGNs (mAGNs) from Ref.~\cite{stecker2019extragalactic}. That mAGN GLF is derived from the core radio luminosity function of Ref.~\cite{yuan2018determining}. The cross-correlation function then becomes

\begin{align}
\Xi_{\rm phys}^{ar}(\theta) &=
A_{\rm mAGN,eff}\, \Xi_{\rm mAGN}^{ar}(\theta, \mu_{\rm mAGN}, M_0, \alpha) \,.
\label{eq:physmdl_2}
\end{align}
For simplicity, we do not include a free parameter for the redshift scaling of the GLF; the model therefore has four free parameters rather than the five used for blazars in the main text. We perform MCMC parameter inference, with the posterior distribution shown in Fig.~\ref{fig:posterior_mAGN_lensing}, \ref{fig:Mhalo_posteriors_lensing_mAGN} and \ref{fig:errorbars_lensing_mAGN}.

\begin{figure*}[!t]
    \centering
        \includegraphics[width=\textwidth]{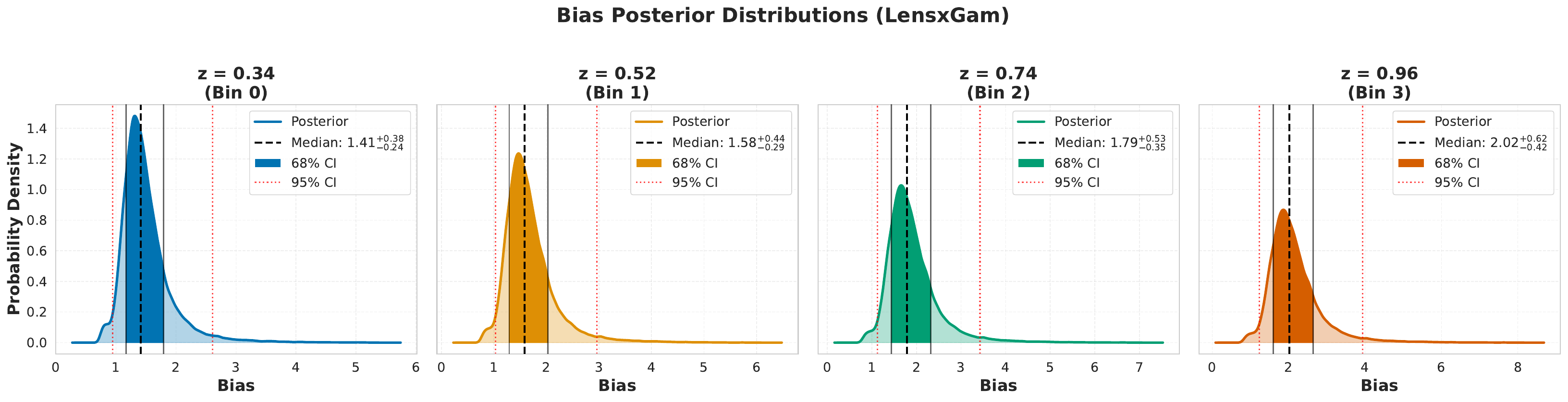}
        \caption{As in Fig. \ref{fig:bias_posteriors_lensing}, but for mAGNs.}
        \label{fig:bias_posteriors_lensing_mAGN}
\end{figure*}

The bias grows across the four DES~Y3 source bins, from $\langle b(z)\rangle = 1.41^{+0.38}_{-0.24}$ at $z=0.34$ to $2.02^{+0.62}_{-0.42}$ at $z=0.96$, about $10$--$20\%$ below the blazar values at every redshift. The host masses are again nearly constant in redshift, with medians $M(z)\sim(0.99\text{-}1.20)\times10^{13}M_\odot$, roughly a factor of two smaller than in the blazar case. As with the blazar case, the growth of $\langle b\rangle$ tracks the evolution of $b_{\rm h}$ rather than an evolving host population. The mAGN and blazar determinations are thus slightly shifted but agree within their $68\%$ credible intervals.
The central result, that moderately biased sources with $\langle b\rangle\gtrsim 2$ at $z\simeq 1$, hosted by $\mathcal{O}(10^{13}M_\odot)$ halos dominate the UGRB signal, is therefore robust to the choice of source population. It is determined by the shape of the cross-correlation rather than by the details of the GLF.

This robustness does not, however, imply that the different populations describe the data equally well. Treating mAGNs as the dominant population leaves the signal-to-noise ratio unchanged relative to the blazar case, but requires a best-fit spectral index of $\mu_{\rm mAGN}=1.80$. The reference mAGN model of Ref.~\cite{stecker2019extragalactic}, by contrast, is a power law with index $\approx 2.3$. The fit therefore requires a population significantly harder than known mAGNs and cannot accommodate the softer reference model. Thus, the mAGN GLF can reproduce the clustering of the signal but not its energy dependence, consistent with the conclusion of Ref.~\cite{thakore2025high} that mAGNs are a subdominant component. This Appendix should therefore be regarded as a cross-check of the dependence of the bias determination on the assumed GLF, rather than as a realistic alternative source model.

\bibliographystyle{apsrev4-2}
\bibliography{bibliography}

\end{document}